\documentclass[conference]{IEEEtran}
\IEEEoverridecommandlockouts
\usepackage{amsmath,amssymb,amsfonts}
\usepackage{array}
\usepackage[caption=false,font=normalsize,labelfont=sf,textfont=sf]{subfig}
\usepackage{textcomp}
\usepackage{stfloats}
\usepackage{verbatim}
\usepackage{graphicx}
\usepackage{cite}
\usepackage{tabularx}
\usepackage{booktabs}

\def\BibTeX{{\rm B\kern-.05em{\sc i\kern-.025em b}\kern-.08em
    T\kern-.1667em\lower.7ex\hbox{E}\kern-.125emX}}

\usepackage{color}

\usepackage{tikz} 
\usepackage[utf8]{inputenc}
\usepackage{pgfplots}
\usepackage{tabu,longtable}
\usepackage{diagbox}
\usepackage{threeparttable}
\usepackage{makecell}
\usepackage{multirow} % enable table multirow
\usepackage[bottom=1.05in,top=0.75in,left=0.7in,right=0.7in]{geometry}
\usepackage{units} 		% use \unit command
\usepackage{bm} 		% for bold symbol
\usepackage{amssymb} 	% for AMS symbols
\newcommand{\TT}{\mathsf{T}}

\newcommand{\bv}{{\bf b}}

\newcommand{\nv}{{\bf n}}

\newcommand{\pv}{{\bf p}}

\newcommand{\rv}{{\bf r}}
\newcommand{\sv}{{\bf s}}

\newcommand{\uv}{{\bf u}}

\newcommand{\Dt}{{\rm D}}

\newcommand{\Lt}{{\rm L}}

\newcommand{\Nt}{{\rm N}}

\newcommand{\muv}{\hbox{\boldmath$\mu$}}

\usepackage{hyperref} 	% for URL
\usepackage{amsmath}

\usepackage{algorithm}
\usepackage{algpseudocode}% http://ctan.org/pkg/algorithmicx
\algtext*{EndWhile}% Remove "end while" text
\algtext*{EndIf}% Remove "end if" text
\algtext*{EndFor}

\makeatletter
\algnewcommand{\LineComment}[1]{\Statex \hskip\ALG@thistlm \(\triangleright\) #1}
\makeatother
\usepackage[printonlyused]{acronym}
\newacro{bs} [BS] {base station}
\newacro{ue} [UE] {user equipment}
\newacro{uav} [UAV] {unmanned aerial vehicle}
\newacro{hap} [HAP] {high altitude platform}
\newacro{ntn} [NTN] {non-terrestrial network}
\newacro{mimo} [MIMO] {multiple-input multiple-output}
\newacro{los} [LoS] {line-of-sight}
\newacro{nlos} [NLoS] {non-line-of-sight}
\newacro{us} [US] {uncorrelated scattering}
\newacro{vmf} [vMF] {von-Mises-Fisher}
\newacro{pdf} [PDF] {probability density function}
\newacro{leo} [LEO] {low Earth orbit}
\newacro{hpbw} [HPBW] {half-power beamwidth}
\usepackage{tikz} 
\usepackage[utf8]{inputenc}
\usepackage{pgfplots} 
\usepackage{tabu,longtable}
\usepackage{diagbox}
\usepackage{threeparttable}
\usepackage{makecell}
\usepackage{bm} % for bold symbol
\usepackage{amssymb} % for AMS symbols
\usepackage{tikz}
\usetikzlibrary{calc,fadings,decorations.pathreplacing}
\usepackage{tikz-3dplot}
\newcommand\pgfmathsinandcos[3]{%
  \pgfmathsetmacro#1{sin(#3)}%
  \pgfmathsetmacro#2{cos(#3)}%
}
\newcommand\LongitudePlane[3][current plane]{%
  \pgfmathsinandcos\sinEl\cosEl{#2} % elevation
  \pgfmathsinandcos\sint\cost{#3} % azimuth
  \tikzset{#1/.style={cm={\cost,\sint*\sinEl,0,\cosEl,(0,0)}}}
}
\newcommand\LatitudePlane[3][current plane]{%
  \pgfmathsinandcos\sinEl\cosEl{#2} % elevation
  \pgfmathsinandcos\sint\cost{#3} % latitude
  \pgfmathsetmacro\yshift{\cosEl*\sint}
  \tikzset{#1/.style={cm={\cost,0,0,\cost*\sinEl,(0,\yshift)}}} %
}
\newcommand\DrawLongitudeCircle[2][1]{
  \LongitudePlane{\angEl}{#2}
  \tikzset{current plane/.prefix style={scale=#1}}
  \pgfmathsetmacro\angVis{atan(sin(#2)*cos(\angEl)/sin(\angEl))} %
  \draw[current plane] (\angVis:1) arc (\angVis:\angVis+180:1);
  \draw[current plane,dashed] (\angVis-180:1) arc (\angVis-180:\angVis:1);
}
\newcommand\DrawLatitudeCircle[2][1]{
  \LatitudePlane{\angEl}{#2}
  \tikzset{current plane/.prefix style={scale=#1}}
  \pgfmathsetmacro\sinVis{sin(#2)/cos(#2)*sin(\angEl)/cos(\angEl)}
  \pgfmathsetmacro\angVis{asin(min(1,max(\sinVis,-1)))}
  \draw[current plane] (\angVis:1) arc (\angVis:-\angVis-180:1);
  \draw[current plane,dashed] (180-\angVis:1) arc (180-\angVis:\angVis:1);
}

\begin{document}

\bstctlcite{IEEEexample:BSTcontrol}
% ------------------------------------------------------------------------------
\title{A Statistical Evaluation of Coherence Time \\for Non-Terrestrial Communications}

\author{
Pinjun Zheng\IEEEauthorrefmark{1}, Anas Chaaban\IEEEauthorrefmark{1}, Md. Jahangir Hossain\IEEEauthorrefmark{1}, Tareq Y. Al-Naffouri\IEEEauthorrefmark{2}\\
\IEEEauthorrefmark{1}\textit{School of Engineering, University of British Columbia, Kelowna, Canada}\\
\IEEEauthorrefmark{2}\textit{CEMSE, King Abdullah University of Science and Technology, Thuwal, Saudi Arabia}\\
(Email: pinjun.zheng@ubc.ca)
}

\markboth{draft}{draft}
\maketitle
% ------------------------------------------------------------------------------

\begin{abstract}
Non-terrestrial networks~(NTNs) present significant challenges for reliable communication due to the dynamic nature of their channels. Studying channel coherence time is crucial, since it directly impacts the design of robust transmission schemes (e.g.,~channel estimation and precoding strategies). This paper evaluates the coherence time of non-terrestrial channels theoretically, revealing that the rapid mobility of non-terrestrial base stations~(BSs) substantially reduces channel coherence time. Our results demonstrate that the presence and enhancement of the line-of-sight~(LoS) channel play a crucial role in extending coherence time, thereby improving the stability of NTN links. Furthermore, unlike terrestrial networks, where beamwidth adjustments can effectively influence coherence time, our findings indicate that in NTNs, receiver beamwidth has a negligible effect under high-speed BS motion. These insights provide valuable guidelines for designing robust transmission schemes and adaptive signal processing techniques in future NTN deployments.
\end{abstract}


\section{Introduction}
 
 Serving as potent supplements to terrestrial networks, non-terrestrial \acp{bs} offer expansive coverage and enhanced efficiency for wireless communication, sensing, and localization services~\cite{3GPPNTN,Giordani2021Non,Saleh2025Integrated,Zheng2023LEO}. Like their terrestrial counterparts, non-terrestrial networks benefit significantly from \ac{mimo} systems and beamforming techniques~\cite{Gesbert2007Shifting,Chen2023Handover,Khammassi2024Precoding,Wang2024Beamforming}, which enhance spectral efficiency and signal robustness. However, unique challenges arise in non-terrestrial environments, including high mobility, frequent handovers, and significant variations in propagation conditions due to altitude and atmospheric effects~\cite{Prol2022Position,Stock2025Survey}. One critical issue is the coherence time of air-to-ground channels, which governs the frequency at which precoders and combiners must be updated to maintain reliable communication. Rapid fluctuations in channel characteristics, influenced by the motion of terrestrial \acp{bs}, Doppler shifts, and environmental factors, further impact the channel coherence time and thus complicate the design of adaptive beamforming strategies~\cite{3GPP38811}. Despite its significance, this issue has not been thoroughly investigated in the literature, highlighting a crucial gap that demands further exploration.

The channel coherence time refers to the duration after which the channel loses its correlation. Typically, the channel coherence time~$T_c$ is approximately the inverse of the Doppler spread~$f_D$, i.e.,~$T_c \simeq 1/f_D$~\cite{Goldsmith2005Wireless}. Non-terrestrial \acp{bs} such as LEO satellites generally experience high-speed mobility, thus inducing significant Doppler frequency shifts in received signals~\cite{3GPPNTN}. For example, an \unit[800]{km}-altitude \ac{leo} satellite at a \unit[45]{$^\circ$} elevation angle can result in an approximately \unit[230]{kHz} Dopper frequency shift in the Ku band~\cite{Shi2024OTFS}. Such an increase in Doppler frequency diminishes the channel coherence time. Apart from that, the coherence time is also related to the beamwidth at the receiver. A narrower beamwidth aids in mitigating destructive multipath components while concurrently accentuating the effect of misalignment. Notably, the results in~\cite{Va2015Basic} and~\cite{Va2017Impact} reveal that by carefully designing beamwidth, the coherence time can be effectively enlarged. 

Nonetheless, all these aforementioned results are based on the terrestrial network without accounting for the mobility of the \ac{bs}, which might not be applicable to non-terrestrial networks. To offer a more accurate and comprehensive assessment of coherence time in air-to-ground channels, this paper extends the analysis in \cite{Va2017Impact} by (i) integrating dual-mobility at both transmitter and receiver ends, (ii) employing a 3D geometric model instead of 2D, and (iii) utilizing accurate calculation of the  Doppler phase shift in integral form~$\phi_\Dt(t) = \int_0^tf_\Dt(\tau)d\tau$ instead of the approximate linear form~$\phi_\Dt(t) = f_\Dt(0)t$. The derivation and results in this paper present a theoretical coherence time evaluation approach for \ac{ntn} channels that offers valuable references and insights for various \ac{ntn}-based applications~\cite{You2020Massive,Abdelsadek2022Distributed,Kim2025Cell,Ma2024Integrated}.

\section{System and Channel Model}\label{sec:SCM}

% pdf/png/jpg figures
\begin{figure}[t]
  \centering
  \includegraphics[width=0.85\linewidth]{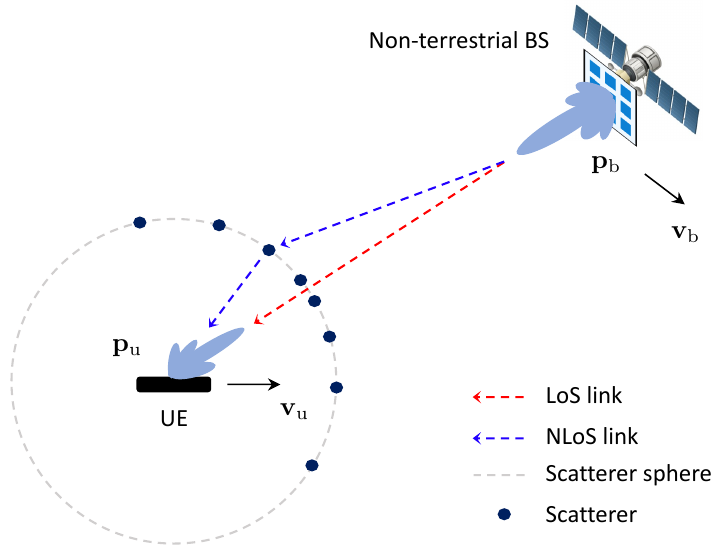}
  \caption{ 
      Illustration of a downlink communication from a mobile non-terrestrial \ac{bs} to a mobile \ac{ue} on the ground. Both \ac{los} and \ac{nlos} links exist and the scatterers are assumed to be distributed on a sphere.
    }
  \label{fig_system}
\end{figure}

We consider a wireless communication system featuring a moving non-terrestrial \ac{bs} and a moving \ac{ue} on the ground, as illustrated in Fig.~\ref{fig_system}. Let~$\pv_\mathrm{b}$,~$\mathbf{v}_\mathrm{b}$,~$\pv_\mathrm{u}$,~$\mathbf{v}_\mathrm{u}\in\mathbb{R}^3$ denote the \ac{bs}'s position, \ac{bs}'s velocity, \ac{ue}'s position, and \ac{ue}'s velocity, respectively. We assume that  velocities~$\mathbf{v}_\mathrm{b}$ and~$\mathbf{v}_\mathrm{u}$ are fixed and hence positions change linearly with time as
\begin{align}
	\pv_\mathrm{b}(t) &= \pv_\mathrm{b}(0) + \mathbf{v}_\mathrm{b}t,\\
	\pv_\mathrm{u}(t) &= \pv_\mathrm{u}(0) + \mathbf{v}_\mathrm{u}t.
\end{align}
By default, we designate $\pv_\mathrm{u}(0)$ as the origin of the coordinate system for all positions.
Since a dominant \ac{los} link is crucial for non-terrestrial communications~\cite{3GPPSAT,Zheng2024LEO}, we examine a scenario with a dominant \ac{los} link and multiple \ac{nlos} links. Hence, the wireless channel can be expressed as~\cite{Zheng2024LEO}
\begin{equation}
	h(t) = \sqrt{\frac{K}{K+1}}h_\Lt(t) + \sqrt{\frac{1}{K+1}}h_\Nt(t),
\end{equation}
where~$K$ is the Rician factor denoting the ratio between the powers of the \ac{los} and \ac{nlos} signals. 

\subsection{The LoS Channel}
The narrowband equivalent lowpass channel impulse response of the \ac{los} path can be written as~\cite{Goldsmith2005Wireless}
\begin{equation}
	h_\Lt(t) = \sqrt{G_\mathrm{b}(t)G_\mathrm{u}(t)}e^{j2\pi(\phi(t)-f_c\xi)},
\end{equation}
where~$G_\mathrm{b}(t)$ and~$G_\mathrm{u}(t)$ respectively denote beam patterns at \ac{bs} and \ac{ue},~$\phi(t)$ is the phase shift due to the Doppler effect,~$f_c$ is the carrier frequency, and~$\xi$ is the channel delay. For both \ac{bs} and \ac{ue}, we utilize a simple beam pattern as~$\cos^q(\theta)$~\cite{Balanis2016Antenna}, where~$q$ is a power factor controls the beam directivity and~$\theta$ denotes the angle between the direction of \ac{los} and the direction of beam major lobe (pointing direction). Then, the beam patterns can be expressed as
\begin{align}
	G_\mathrm{b}(t) &= \bigg(\frac{\big(\pv_\mathrm{u}(t)-\pv_\mathrm{b}(t)\big)^\TT\bv }{\|\pv_\mathrm{u}(t)-\pv_\mathrm{b}(t)\|}\bigg)^{q_\mathrm{b}},\label{eq:Gb}\\
	G_\mathrm{u}(t) &= \bigg(\frac{\big(\pv_\mathrm{b}(t)-\pv_\mathrm{u}(t)\big)^\TT\uv }{\|\pv_\mathrm{b}(t)-\pv_\mathrm{u}(t)\|}\bigg)^{q_\mathrm{u}},\label{eq:Gu}
\end{align}
where~$\bv\in\mathbb{R}^3$ and~$\uv\in\mathbb{R}^3$ respectively denote the unit directional vectors of the major lobes at the \ac{bs} and \ac{ue}, and~$q_\mathrm{b}$ and~$q_\mathrm{u}$ respectively stand for the directivity of the beams at the \ac{bs} and \ac{ue}. The dependence of the beam pattern on time~$t$ signifies the beam misalignment. In addition, the power factor~$q$ can be related to the \ac{hpbw} (denoted as~$\psi$) by $\psi = 2 \arccos(2^{-{1}/{q}})$.

Since the Doppler frequency at time~$t$ is given by~\cite{Goldsmith2005Wireless}
\begin{equation}
	f_\Dt(t)=\frac{(\mathbf{v}_\mathrm{u}-\mathbf{v}_\mathrm{b})^\TT(\pv_\mathrm{b}(t)-\pv_\mathrm{u}(t))}{\lambda\|\pv_\mathrm{b}(t)-\pv_\mathrm{u}(t)\|},
\end{equation}
where $\lambda$ denotes signal wavelength, the Doppler shift~$\phi(t)$ can be obtained as
\begin{align}
	&\phi(t) = \int_0^t\frac{(\mathbf{v}_\mathrm{u}-\mathbf{v}_\mathrm{b})^\TT\big(\pv_\mathrm{b}(\tau)-\pv_\mathrm{u}(\tau)\big)}{\lambda\|\pv_\mathrm{b}(\tau)-\pv_\mathrm{u}(\tau)\|} d\tau, \\
	&= -\frac{1}{\lambda} \int_0^t \frac{\mathbf{v}^\TT(\pv_0+\mathbf{v}\tau)}{\|\pv_0+\mathbf{v}\tau\|} d\tau = -\frac{1}{\lambda}\big(\|\pv_0+\mathbf{v}t\|-\|\pv_0\|\big), \notag
\end{align}
where~$\pv_0=\pv_\mathrm{b}(0)-\pv_\mathrm{u}(0)$ and~$\mathbf{v}=\mathbf{v}_\mathrm{b}-\mathbf{v}_\mathrm{u}$.

\subsection{The NLoS Channel}
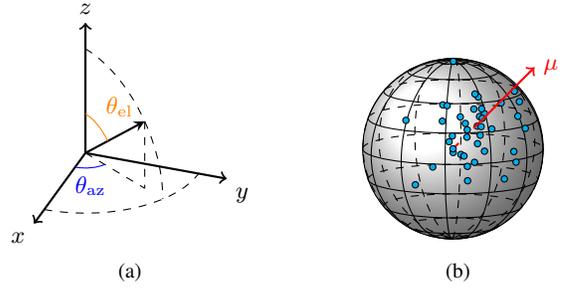
\begin{figure}[t]
\centering
\small

  \begin{minipage}[t]{0.48\linewidth}
  \centering	
	%set the plot display orientation
	%synatax: \tdplotsetdisplay{\theta_d}{\phi_d}
	\tdplotsetmaincoords{60}{110}
	%define polar coordinates for some vector
	%TODO: look into using 3d spherical coordinate system
	\pgfmathsetmacro{\rvec}{.8}
	\pgfmathsetmacro{\thetavec}{50}
	\pgfmathsetmacro{\phivec}{60}
	%start tikz picture, and use the tdplot_main_coords style to implement the display 
	%coordinate transformation provided by 3dplot
	\begin{tikzpicture}[scale=2,tdplot_main_coords]
	%set up some coordinates 
	%-----------------------
	\coordinate (O) at (0,0,0);
	%determine a coordinate (P) using (r,\theta,\phi) coordinates.  This command
	%also determines (Pxy), (Pxz), and (Pyz): the xy-, xz-, and yz-projections
	%of the point (P).
	%syntax: \tdplotsetcoord{Coordinate name without parentheses}{r}{\theta}{\phi}
	\tdplotsetcoord{P}{\rvec}{\thetavec}{\phivec}
	%draw figure contents
	%--------------------
	%draw the main coordinate system axes
	\draw[thick,->] (0,0,0) -- (1,0,0) node[anchor=north east]{$x$};
	\draw[thick,->] (0,0,0) -- (0,1,0) node[anchor=north west]{$y$};
	\draw[thick,->] (0,0,0) -- (0,0,1) node[anchor=south]{$z$};
	%draw a vector from origin to point (P) 
	\draw[-stealth,color=black,thick] (O) -- (P);
	%draw projection on xy plane, and a connecting line
	\draw[dashed, color=black] (O) -- (Pxy);
	\draw[dashed, color=black] (P) -- (Pxy);
	%draw the angle \phi, and label it
	%syntax: \tdplotdrawarc[coordinate frame, draw options]{center point}{r}{angle}{label options}{label}
	\tdplotdrawarc[color=blue]{(O)}{0.2}{0}{\phivec}{anchor=north}{$\theta_\mathrm{az}$}
	%set the rotated coordinate system so the x'-y' plane lies within the
	%"theta plane" of the main coordinate system
	%syntax: \tdplotsetthetaplanecoords{\phi}
	\tdplotsetthetaplanecoords{\phivec}
	%draw theta arc and label, using rotated coordinate system
	\tdplotdrawarc[tdplot_rotated_coords,color=orange]{(0,0,0)}{0.3}{0}{\thetavec}{anchor=south west}{$\theta_\mathrm{el}$}
	%draw some dashed arcs, demonstrating direct arc drawing
	\draw[dashed,tdplot_rotated_coords] (\rvec,0,0) arc (0:90:\rvec);
	\draw[dashed] (\rvec,0,0) arc (0:90:\rvec);
  \end{tikzpicture}
  \centerline{\footnotesize(a)}
\end{minipage}
\begin{minipage}[t]{0.48\linewidth}
  \centering 
  \begin{tikzpicture}[scale=0.6] % "THE GLOBE" showcase
	\def\R{2} % sphere radius
	\def\angEl{10} % elevation angle
	\filldraw[ball color=white] (0,0) circle (\R);
	\foreach \t in {-80,-60,...,80} { \DrawLatitudeCircle[\R]{\t} }
	\foreach \t in {-5,-35,...,-175} { \DrawLongitudeCircle[\R]{\t} }
	\def\points{%
(1.3210,    0.9015,    1.2010),
(0.9813,    0.2732,    1.7212),
(1.0572,    1.1846,    1.2162),
(0.9183,    1.5105,    0.9354),
(1.0597,    1.0225,    1.3534),
(0.8899,    0.9950,    1.4893),
(1.1450,    0.8426,    1.4068),
(0.9098,    0.8588,    1.5604),
(0.6250,    0.9242,    1.6599),
(-0.0642,   -0.0283,   1.9988),
(0.8426,    0.5407,    1.7314),
(0.6890,    0.6022,    1.7784),
(1.6617,    0.9479,    0.5832),
(1.1061,    0.3287,    1.6336),
(1.1300,    0.9980,    1.3142),
(0.3612,    1.2105,    1.5506),
(0.3009,    0.3456,    1.9468),
(0.9035,    0.4990,    1.7131),
(1.2620,    0.5109,    1.4650),
(-0.388,    1.2863,    1.7012),
(1.1225,    0.6922,    1.5036),
(1.4455,    1.3645,    0.2208),
(1.5932,    1.1669,    0.3165),
(0.8401,    1.5672,    0.9154),
(1.1293,    1.2111,    1.1216),
(1.0541,    1.3049,    1.0891),
(0.5801,    0.8224,    1.7283),
(0.2814,    1.4802,    1.3152),
(1.7611,    0.0687,    0.9454),
(0.9920,   -0.0400,    1.7362),
(0.8271,    1.1161,    1.4388),
(0.9160,    1.3331,    1.1764),
(1.1138,    1.1408,    1.2075),
(1.5954,   -0.2102,    1.1877),
(1.8093,    0.6606,    0.5386),
(1.7211,    0.3130,    0.9694),
(0.7014,    1.2450,    1.3993),
(1.3436,    1.2955,    0.7185),
(0.3764,    1.4617,    1.3122),
(0.0622,  1.9939,  0.1425),
}
	\draw[red,dashed,thick] (0,0,0) -- (1,1,1.4142);
	\foreach \p in \points {
    	\draw[black,fill=cyan] \p circle (0.07);
	}
	\draw[red,->,thick] (1,1,1.4142) -- (4,4,5.6968) node[right] {$\bf{\mu}$};
  \end{tikzpicture}
  \centerline{\footnotesize (b)}
\end{minipage}

\caption{
    Geometric setups. (a) The definition of the azimuth and elevation components of a solid angle. (b) A realization of random scatterers conformed to 3D \ac{vmf} distribution.
  }
  \label{Fig_geo}
\end{figure}

Analogous to~\cite{Va2017Impact}, we assume that the scatterers are distributed on a 3D sphere~$\mathbb{S}^2$ with radius~$R$ centered around the \ac{ue}, and that the reflected signals are uncorrelated with each other (known as the \ac{us} assumption~\cite{Goldsmith2005Wireless}). Then, the \ac{nlos} channel impulse response can be characterized as~\cite{Va2015Basic,Va2017Impact}
\begin{multline}
	h_\Nt(t) = \int_{\mathbb{S}^2}  \sqrt{p(\Omega)G_\mathrm{b}(t, \Omega)G_\mathrm{u}(t, \Omega)} \\ \times e^{j2\pi(\varphi(t,\Omega) + \gamma(\Omega) - f_c\zeta(\Omega))} d\Omega,
\end{multline}
where~$p(\Omega)$ is the \ac{pdf} of the scatterers' distribution on the sphere,~$\varphi(t,\Omega)$ is the Doppler shift in \ac{nlos} links,~$\gamma(\Omega)\sim U[0,1)$ is a uniformly distributed random phase shift associated with each scatterer, and~$\zeta(\Omega)$ denotes the \ac{nlos} channel delay. Here,~$\mathbb{S}^2\subset\mathbb{R}^3$ represents the considered 3D sphere and~$\Omega$ is the solid angle corresponding to an infinitesimal area on the sphere. Typically, each solid angle~$\Omega$ can be uniquely characterized using an azimuth angle~$\theta_\mathrm{az}$ and an elevation angle~$\theta_\mathrm{el}$, i.e.,~$\Omega=\{\theta_\mathrm{az},\theta_\mathrm{el}\}$, as shown in Fig.~\ref{Fig_geo}-(a). The term~$p(\Omega)G_\mathrm{b}(t,\Omega)G_\mathrm{u}(t,\Omega)$ is also known as the effective power angular spectrum~\cite{Va2015Basic}. 

In this work, we adopt the 3D \ac{vmf} distribution to describe the distribution of scatterers over the sphere. Specifically, a 3D \ac{vmf} distribution has the   \ac{pdf}~\cite{Straub2017Bayesian}
\begin{equation}
	p(\Omega;\muv,\rho) = \frac{\rho e^{\rho\bm{\mu}^\TT\nv(\Omega)}}{2\pi(e^{\rho}-e^{-\rho})},
\end{equation}
where~$\muv\in\mathbb{R}^3$ is a unit-norm vector indicating the mean direction,~$\rho>0$ is the concentration parameter, and~$\nv(\Omega)$ is the directional vector corresponding to the solid angle~$\Omega$, i.e.,~$\nv(\Omega)=[\cos(\theta_\mathrm{az})\sin(\theta_\mathrm{el}),\sin(\theta_\mathrm{az})\sin(\theta_\mathrm{el}),\cos(\theta_\mathrm{el})]^\TT$. A higher value of $\rho$ indicates that the scatterers are more concentrated around the mean direction $\muv$. Figure~\ref{Fig_geo}-(b) visualizes a realization of a set of \ac{vmf}-distributed scatterers on a sphere. 

The beam patterns~$G_\mathrm{b}(t,\Omega)$ and~$G_\mathrm{u}(t,\Omega)$ can be obtained according to~\eqref{eq:Gb} and~\eqref{eq:Gu} by replacing~$\pv_\mathrm{u}(t)$ and~$\pv_\mathrm{b}(t)$ with~$\pv(\Omega)=R\nv(\Omega)$. 
Since the Doppler frequency of the \ac{nlos} signal corresponding to the solid angle~$\Omega$ is given by
\begin{equation}
f_\mathrm{D}(t,\Omega)=\frac{\mathbf{v}_\mathrm{b}^\TT (\pv(\Omega)-\pv_\mathrm{b}(t))}{\lambda\|\pv(\Omega)-\pv_\mathrm{b}(t)\|} + \frac{\mathbf{v}_\mathrm{u}^\TT (\pv(\Omega)-\pv_\mathrm{u}(t))}{\lambda\|\pv(\Omega)-\pv_\mathrm{u}(t)\|},
\end{equation}
the Doppler shift~$\varphi(t,\Omega)$ can be obtained as
\begin{align}
	&\varphi(t,\Omega) =\! \int_0^t \frac{\mathbf{v}_\mathrm{b}^\TT (\rv(\Omega)-\mathbf{v}_\mathrm{b}\tau)}{\lambda\|\rv(\Omega)-\mathbf{v}_\mathrm{b}\tau\|} d\tau\! +\! \int_0^t \frac{\mathbf{v}_\mathrm{u}^\TT (\sv(\Omega)-\mathbf{v}_\mathrm{u}\tau)}{\lambda\|\sv(\Omega)-\mathbf{v}_\mathrm{u}\tau\|} d\tau, \notag\\
	&=\!-\frac{1}{\lambda}\big(\|\rv(\Omega)\!-\!\mathbf{v}_\mathrm{b}t\|\!-\!\|\rv(\Omega)\| \!+\! \|\sv(\Omega)\!-\!\mathbf{v}_\mathrm{u}t\|\!-\!\|\sv(\Omega)\|\big),\!
\end{align}
where~$\rv(\Omega)=\pv(\Omega)-\pv_\mathrm{b}(0)$ and~$\sv(\Omega)=\pv(\Omega)-\pv_\mathrm{u}(0)$.

\section{Channel Autocorrelation and Coherence Time Derivation}

The channel coherence time~$T_c$ is defined as a time after which the channel autocorrelation decreases to a predefined threshold~$\epsilon$~\cite{Goldsmith2005Wireless,Va2015Basic,Va2017Impact}. The channel decorrelation results from two effects: (i) the superimposed distortion effect of Doppler shifts and random phases caused by different multipath scatterers and (ii) the beam misalignment due to the \ac{bs}'s and/or \ac{ue}'s movements. The model presented in Section~\ref{sec:SCM} has taken these two effects into account. 

Although \ac{nlos} channels are commonly assumed as wide-sense stationary~\cite{Goldsmith2005Wireless}, this assumption does not hold for the \ac{los} channel. Therefore, we keep the time variable~$t$ and write the channel autocorrelation as 
\begin{align}
	&A_h(t,\tau) = \mathbb{E}[h(t)h^*(t+\tau)], \label{eq:Ah}\\
	&= \frac{K}{K+1}\mathbb{E}[h_\Lt(t)h_\Lt^*(t+\tau)] + \frac{\sqrt{K}}{K+1}\mathbb{E}[h_\Lt(t)h_\Nt^*(t+\tau)]\notag\\ 
	&\quad + \frac{\sqrt{K}}{K+1}\mathbb{E}[h_\Nt(t)h_\Lt^*(t+\tau)] + \frac{1}{K+1}\mathbb{E}[h_\Nt(t)h_\Nt^*(t+\tau)].\notag
\end{align}
Then, these four terms are derived individually in~\eqref{eq:term1}--\eqref{eq:term4} at the top of the next page. Note that the \ac{los} channel~$h_\Lt(t)$ is a deterministic function of~$t$ while the \ac{nlos} channel~$h_\Nt(t)$ is a stochastic process. Step~(a) in~\eqref{eq:term2} and step~(b) in~\eqref{eq:term3} follows since $\gamma(\Omega)\sim U[0,1)$, leading to $\mathbb{E}[e^{-j2\pi\gamma(\Omega)}]=0$ and so 
\begin{multline}
	\mathbb{E}[h_\Nt(t)]=\int_{\mathbb{S}^2}  {\sqrt{p(  \Omega  )G_\mathrm{b}  (  t,  \Omega  )G_\mathrm{u}  (  t,  \Omega  )}} e^{j2\pi(\varphi(t,\Omega)-f_c\zeta(\Omega))} \\ \times \mathbb{E}\big[e^{j2\pi\gamma(\Omega)}\big] d\Omega =0.
\end{multline}
Moreover, step~(c) in~\eqref{eq:term4} follows from the \ac{us} assumption. Specifically, since the random phase shifts caused by different scatterers are uncorrelated, we have 
\begin{align}
	&\mathbb{E} \big[e^{j2\pi(\gamma(\Omega_1)-\gamma(\Omega_2))}\big] \\
& = \begin{cases}
\mathbb{E} \big[e^{j2\pi(\gamma(\Omega_1))}\big] \mathbb{E} \big[e^{-j2\pi(\gamma(\Omega_2))}\big]=0, & \text{if } \Omega_1\neq\Omega_2,\\
1, & \text{if } \Omega_1=\Omega_2, \notag
\end{cases}
\end{align}
from which~\eqref{eq:term4} follows. It is not possible to further simplify~\eqref{eq:term4}, but we can evaluate it numerically to gain insight into the \ac{ntn} channel coherence time. Specifically, the integral over~$\mathbb{S}^2$ can be implemented by integrating the objective function over the azimuth and elevation components~\cite{Balanis2016Antenna}
\begin{equation}
	\int_{\mathbb{S}^2} f(\Omega) d\Omega = \int_0^{2\pi}\int_0^{\pi} f(\theta_\mathrm{az},\theta_\mathrm{el})\sin(\theta_\mathrm{el}) d\theta_\mathrm{el} d\theta_\mathrm{az}.
\end{equation}

\begin{figure*}
\footnotesize
\begin{align}
	&\mathbb{E}[h_\Lt(t)h_\Lt^*(t+\tau)] = \sqrt{G_\mathrm{b}(t)G_\mathrm{b}(t+\tau)G_\mathrm{u}(t)G_\mathrm{u}(t+\tau)}e^{j2\pi(\phi(t)-\phi(t+\tau))}, \label{eq:term1} \\
	&\mathbb{E}[h_\Lt(t)h_\Nt^*(t+\tau)] = h_\Lt(t)\mathbb{E}[h_\Nt^*(t+\tau)] \stackrel{(a)}{=} 0, \label{eq:term2} \\
	&\mathbb{E}[h_\Nt(t)h_\Lt^*(t+\tau)] = {h_\Lt^*(t+\tau)} \mathbb{E}[h_\Nt(t)] \stackrel{(b)}{=} 0, \label{eq:term3}\\
	&\mathbb{E}[h_\Nt(t)h_\Nt^*(t+\tau)] \notag\\
	& = \!\mathbb{E} \Big[ \!\int_{\mathbb{S}^2}\!\int_{\mathbb{S}^2}\! \sqrt{\!p(\Omega_1)p(\Omega_2) G_\mathrm{b}(t,\Omega_1)G_\mathrm{b}(t\!+\!\tau,\Omega_2)G_\mathrm{u}(t,\Omega_1)G_\mathrm{u}(t\!+\!\tau,\Omega_2)}e^{j2\pi(\varphi(t,\Omega_1)-\varphi(t+\tau,\Omega_2))} e^{j2\pi(\gamma(\Omega_1)-\gamma(\Omega_2))} e^{j2\pi f_c(\zeta(\Omega_2)-\zeta(\Omega_1))} d\Omega_1 d\Omega_2\Big],\notag \\
	&=  \!\int_{\mathbb{S}^2}\!\int_{\mathbb{S}^2}\! \sqrt{\!p(\Omega_1)p(\Omega_2)G_\mathrm{b}(t,\Omega_1)G_\mathrm{b}(t\!+\!\tau,\Omega_2)G_\mathrm{u}(t,\Omega_1)G_\mathrm{u}(t+\tau,\Omega_2)}e^{j2\pi(\varphi(t,\Omega_1)-\varphi(t\!+\!\tau,\Omega_2))}  \mathbb{E} \Big[e^{j2\pi(\gamma(\Omega_1)-\gamma(\Omega_2))}\Big] e^{j2\pi f_c(\zeta(\Omega_2)-\zeta(\Omega_1))} d\Omega_1 d\Omega_2,\notag\\
	&\stackrel{(c)}{=}  \int_{\mathbb{S}^2} \sqrt{p^2(\Omega) G_\mathrm{b}(t,\Omega)G_\mathrm{b}(t+\tau,\Omega)G_\mathrm{u}(t,\Omega)G_\mathrm{u}(t+\tau,\Omega)} e^{j2\pi(\varphi(t,\Omega)-\varphi(t+\tau,\Omega))} d\Omega. \label{eq:term4}
\end{align}
\hrulefill
\end{figure*}

A normalization step is further applied to~$A_h(t,\tau)$, so that we obtain $\bar{A}_h(t,\tau) = {A_h(t,\tau)}/{A_h(t,0)}$.
Subsequently, we can define the channel coherence time~$T_c$ as
\begin{align}\label{eq:Tc}
	T_c(t) = \min\ \{\ \tau\ |\ |\bar{A}_h(t,\tau)| < \epsilon\ \},\quad 0<\epsilon<1.
\end{align}
Here, the coherence time~$T_c$ is a function of~$t$ since the \ac{los} channel is not statistically stationary.

\section{Numerical Results}

\subsection{Simulation Setup}

This section conducts numerical simulations. As an example, the \ac{bs} is assumed to be deployed in a \ac{leo} satellite~\cite{Su2019Broadband}, performing downlink communications to a ground user at a carrier frequency~$f_c=\unit[28]{GHz}$. By default, we set~$\pv_\mathrm{b}(0)=[-1,0,500]^\TT\ \mathrm{{km}}$, $\mathbf{v}_\mathrm{b}=[7,0,0]^\TT\ \mathrm{{km/s}}$, $\pv_\mathrm{u}(0)=[0,0,0]^\TT\ \mathrm{m}$, $\mathbf{v}_\mathrm{u}=[0,2,0]^\TT\ \mathrm{m/s}$. The radius of the scatterer sphere is set as~$R=\unit[1000]{\lambda}$, where~$\lambda$ denotes the signal wavelength. We set the concentration of the scatterers as~$\rho=30$, and the mean direction~$\muv$ as the \ac{los} direction at~$t=0$. Additionally, the \acp{hpbw}~$\psi$ at the \ac{bs} and \ac{ue} are~$\unit[2]{^\circ}$ and $\unit[20]{^\circ}$, respectively. In the evaluations of the autocorrelation~$\bar{A}_h(t,\tau)$, we pick~$t=0$ thus observing over~$\tau$ to determine the coherence time~$T_c(0)$ according to~\eqref{eq:Tc}.

\begin{figure}[t]
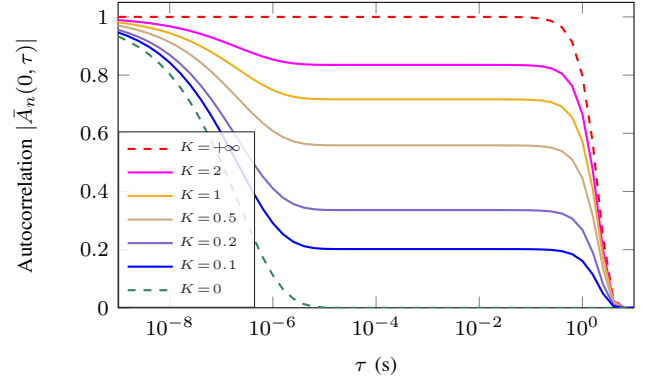

  \centering
  \include{figs/sim1.tex}
  \vspace{-2.5em}
  \caption{ 
    The evaluation of $|\bar{A}_h(0,\tau)|$ vs.~$\tau$ over Rician factor~$K=\{0,0.1,0.2,0.5,1,2,+\infty\}$, where~$K=0$ and~$K=+\infty$ correspond to the \ac{nlos}-only and \ac{los}-only cases, respectively.
    }
  \label{fig_sim1}
\end{figure}

\subsection{Evaluation Results}

Figure~\ref{fig_sim1} plots~$|\bar{A}_h(0,\tau)|$ vs.~$\tau$ across different Rician factors~$K$. Observe that a stronger LoS component (higher~$K$) can effectively delay the decorrelation of the channel. In particular, the coherence time of the \ac{los} component ($K=+\infty$) can be significantly longer than that of the \ac{nlos} component ($K=0$) by up to six orders of magnitude. This underscores the paramount importance of the \ac{los} channel in non-terrestrial communications.  For cases~$0<K<+\infty$ where both the \ac{los} and \ac{nlos} components exist, the channel autocorrelation initially decreases with $\tau$ due to the decorrelation of the \ac{nlos} component. It then stabilizes once the \ac{nlos} component fully decorrelates and eventually diminishes to zero when the \ac{los} component also decorrelates completely. During this process, a higher $K$ value can aid in the preservation of channel correlation, thus sustaining a higher autocorrelation level for a larger interval of $\tau$. 

\begin{figure}[t]
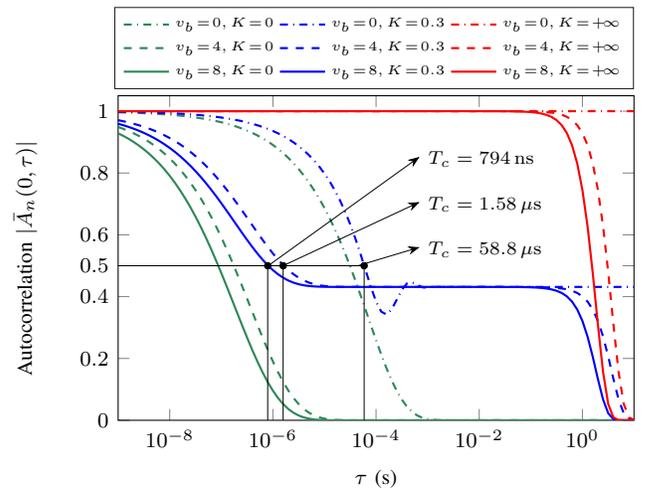

  \centering
  \include{figs/sim2.tex}
  \vspace{-2.5em}
  \caption{ 
    The evaluation of $|\bar{A}_h(0,\tau)|$ vs.~$\tau$ with different speeds of the \ac{bs}. We set~$\mathbf{v}_\mathrm{b}=[v_b,0,0]^\TT$ and test over~$v_b = \{0,4,8\}~\mathrm{km/s}$. The evaluation is also performed over~$K=\{0,0.3,+\infty\}$.
    }
  \label{fig_sim2}
\end{figure}

To investigate the impact of the \ac{bs}'s mobility, Fig.~\ref{fig_sim2} illustrates~$|\bar{A}_h(0,\tau)|$ vs.~$\tau$ when the \ac{leo} satellite moves with different speeds~$v_b=\{0,4,8\}\ \mathrm{km/s}$ with~$\mathbf{v}_\mathrm{b}=[v_b,0,0]^\TT$. Benchmarked against the $v_b=0$ case, the mobility of the \ac{leo} satellite substantially accelerates the channel decorrelation and thus reduces the coherence time. Figure~\ref{fig_sim2} shows an example with~$\epsilon=0.5$ and~$K=0.3$. When~$v_b=0$, the coherence time $T_c$ is approximately~$\unit[58.8]{\mu s}$, which is aligned with the results in~\cite{Va2015Basic,Va2017Impact}. However, when we enlarge the \ac{bs}'s speed to~\unit[4]{km/s} and~\unit[8]{km/s},\footnote{The typical orbit speed of \ac{leo} satellites is~\unit[5--10]{km/s}~\cite{Su2019Broadband}. However, the velocity~$\mathbf{v}_\mathrm{b}$ defined in this paper is the relative velocity of the satellite w.r.t. the ground user. Determining it requires further consideration of the Earth's rotation, which can vary depending on the satellite orbit and position.} the coherence times are reduced to~$\unit[1.58]{\mu s}$ and~$\unit[0.794]{\mu s}$, respectively. This reveals that a high-speed moving \ac{bs} would result in a significant reduction in coherence time, which can even make beamforming impossible considering a typical mmWave OFDM symbol duration of several microseconds~\cite{Mezzavilla2018End}. Based on Fig.~\ref{fig_sim1} and Fig.~\ref{fig_sim2}, enhancing the \ac{los} link could be an effective solution. 

% tikz figures
\begin{figure}[t]
\centering
  \begin{minipage}[b]{1\linewidth}
    \centering
\definecolor{mycolor1}{rgb}{1.00000,0.00000,1.00000}%
\definecolor{mycolor2}{rgb}{0.00000,1.00000,1.00000}%
\begin{tikzpicture}

\begin{axis}[%
width=2.7in,
height=1.8in,
at={(0in,0in)},
scale only axis,
xmode=log,
xmin=8e-10,
xmax=0.02,
xminorticks=true,
xticklabel style = {font=\color{white!15!black},font=\footnotesize},
xlabel style={font=\color{white!15!black},font=\footnotesize},
xlabel={$\tau$ (s)},
ymin=-0.02,
ymax=1.02,
yticklabel style = {font=\color{white!15!black},font=\footnotesize},
ylabel style={font=\color{white!15!black},font=\footnotesize},
ylabel={Autocorrelation $|\bar{A}_n(0,\tau)|$},
xmajorgrids,
xminorgrids,
ymajorgrids,
yminorgrids,
axis background/.style={fill=white},
legend style={at={(1,1)}, anchor=north east, legend cell align=left, align=left, draw=white!15!black, fill opacity=0.85, font=\tiny}
]
\addplot [color=red, mark=o, mark options={solid, red}, line width=0.8pt, mark size=2.1pt]
  table[row sep=crcr]{%
0	1\\
1e-09	0.998716467262107\\
3.16227766016838e-09	0.997718662791519\\
1e-08	0.995946730526946\\
3.16227766016838e-08	0.992803420430336\\
1e-07	0.987237743266096\\
3.16227766016838e-07	0.977414713967222\\
1e-06	0.959386019549443\\
3.16227766016838e-06	0.928837663043648\\
1e-05	0.876561293394399\\
3.16227766016838e-05	0.78928483950296\\
0.0001	0.649618588308103\\
0.000316227766016838	0.445350932835489\\
0.001	0.206254317363206\\
0.00316227766016838	0.0407501970286859\\
0.01	0.000792357562135867\\
};
\addlegendentry{$\psi\!=\!2^\circ,v_b\!=\!0$}

\addplot [color=mycolor1, mark=triangle, mark options={solid, mycolor1}, line width=0.8pt, mark size=2.5pt]
  table[row sep=crcr]{%
0	1\\
1e-09	0.997985167206073\\
3.16227766016838e-09	0.99641986368219\\
1e-08	0.993527794739577\\
3.16227766016838e-08	0.988519239151981\\
1e-07	0.979673231191199\\
3.16227766016838e-07	0.964128524668245\\
1e-06	0.937044576178954\\
3.16227766016838e-06	0.890509214969828\\
1e-05	0.812313384194119\\
3.16227766016838e-05	0.685704988632215\\
0.0001	0.495482781787072\\
0.000316227766016838	0.257381398181301\\
0.001	0.0651724255783361\\
0.00316227766016838	0.00347387385933238\\
0.01	0.000182900195592126\\
};
\addlegendentry{$\psi\!=\!5^\circ,v_b\!=\!0$}

\addplot [color=blue, mark=triangle, mark options={solid, rotate=180, blue}, line width=0.8pt, mark size=2.5pt]
  table[row sep=crcr]{%
0	1\\
1e-09	0.99718456923498\\
3.16227766016838e-09	0.994998827775546\\
1e-08	0.991123659878216\\
3.16227766016838e-08	0.984268990514047\\
1e-07	0.972192099017747\\
3.16227766016838e-07	0.951058112475853\\
1e-06	0.914487927752601\\
3.16227766016838e-06	0.852341056825461\\
1e-05	0.749764377729812\\
3.16227766016838e-05	0.589195770512198\\
0.0001	0.366493918069499\\
0.000316227766016838	0.13761741031989\\
0.001	0.017280905605787\\
0.00316227766016838	0.00020611480063802\\
0.01	5.24549266813265e-05\\
};
\addlegendentry{$\psi\!=\!10^\circ,v_b\!=\!0$}

\addplot [color=mycolor2, mark=square, mark options={solid, mycolor2}, line width=0.8pt, mark size=2pt]
  table[row sep=crcr]{%
0	1\\
1e-09	0.996365817155269\\
3.16227766016838e-09	0.993546477495607\\
1e-08	0.988552299355923\\
3.16227766016838e-08	0.979731531989041\\
1e-07	0.964230760605416\\
3.16227766016838e-07	0.937222797315313\\
1e-06	0.89081578472058\\
3.16227766016838e-06	0.812835818321086\\
1e-05	0.686587127381718\\
3.16227766016838e-05	0.496922150108309\\
0.0001	0.259364691782335\\
0.000316227766016838	0.0667695258463574\\
0.001	0.00379616184648495\\
0.00316227766016838	9.77942662474532e-06\\
0.01	1.88410478245731e-05\\
};
\addlegendentry{$\psi\!=\!20^\circ,v_b\!=\!0$}

\end{axis}

\begin{axis}[%
width=2.7in,
height=1.8in,
at={(0in,0in)},
scale only axis,
xmode=log,
xmin=8e-10,
xmax=0.02,
ymin=-0.02,
ymax=1.02,
xtick={\empty},
ytick={\empty},
legend style={at={(0,0)}, anchor=south west, legend cell align=left, align=left, draw=white!15!black, fill opacity=0.85, font=\tiny}
]
\addplot [color=red, dashed, mark=o, mark options={solid, red}, line width=0.8pt, mark size=3pt]
  table[row sep=crcr]{%
0	1\\
1e-09	0.932967720647907\\
2.51188643150958e-09	0.895856700252963\\
6.30957344480194e-09	0.840049236759804\\
1.58489319246111e-08	0.758634577105304\\
3.98107170553497e-08	0.645451362218126\\
1e-07	0.499637437969462\\
2.51188643150958e-07	0.332966327185935\\
6.30957344480193e-07	0.175007451195271\\
1.58489319246111e-06	0.0631274187231601\\
6.30957344480193e-06	0.00398274355177245\\
3.98107170553497e-05	9.58629926421543e-07\\
0.000251188643150958	7.90108318720738e-16\\
0.00158489319246111	1.45809435904857e-38\\
0.01	2.47394473201584e-94\\
};
\addlegendentry{$\psi\!=\!2^\circ,v_b\!=\!7$}

\addplot [color=mycolor1, dashed, mark=triangle, mark options={solid, mycolor1}, line width=0.8pt, mark size=3pt]
  table[row sep=crcr]{%
0	1\\
1e-09	0.932967800981575\\
2.51188643150958e-09	0.895856924278155\\
6.30957344480194e-09	0.840049733271027\\
1.58489319246111e-08	0.75863472237582\\
3.98107170553497e-08	0.645452851631624\\
1e-07	0.499639181998099\\
2.51188643150958e-07	0.332968296051634\\
6.30957344480193e-07	0.175009765465525\\
1.58489319246111e-06	0.0631453135352781\\
6.30957344480193e-06	0.00403330878306494\\
3.98107170553497e-05	9.71281917841052e-07\\
0.000251188643150958	8.02950436657611e-16\\
0.00158489319246111	1.50995718535402e-38\\
0.01	2.89866316958721e-94\\
};
\addlegendentry{$\psi\!=\!5^\circ,v_b\!=\!7$}

\addplot [color=blue, dashed, mark=triangle, mark options={solid, rotate=180, blue}, line width=0.8pt, mark size=3pt]
  table[row sep=crcr]{%
0	1\\
1e-09	0.932970869348441\\
2.51188643150958e-09	0.89586098383903\\
6.30957344480194e-09	0.840051487117041\\
1.58489319246111e-08	0.758639617384825\\
3.98107170553497e-08	0.645456625848073\\
1e-07	0.499644389438873\\
2.51188643150958e-07	0.332975022437695\\
6.30957344480193e-07	0.175016986771684\\
1.58489319246111e-06	0.0631509378762959\\
6.30957344480193e-06	0.00404149106093541\\
3.98107170553497e-05	9.74668253569634e-07\\
0.000251188643150958	8.12969052662135e-16\\
0.00158489319246111	1.61690856074234e-38\\
0.01	4.47459881431903e-94\\
};
\addlegendentry{$\psi\!=\!10^\circ,v_b\!=\!7$}

\addplot [color=mycolor2, dashed, mark=square, mark options={solid, mycolor2}, line width=0.8pt, mark size=2.2pt]
  table[row sep=crcr]{%
0	1\\
1e-09	0.932973892666838\\
2.51188643150958e-09	0.895864129188421\\
6.30957344480194e-09	0.840055585039947\\
1.58489319246111e-08	0.758644349561907\\
3.98107170553497e-08	0.645465303302579\\
1e-07	0.499656729253026\\
2.51188643150958e-07	0.332991458708686\\
6.30957344480193e-07	0.175034719509581\\
1.58489319246111e-06	0.0631648208263572\\
6.30957344480193e-06	0.00404455919562441\\
3.98107170553497e-05	9.78918228175366e-07\\
0.000251188643150958	8.3472143797321e-16\\
0.00158489319246111	1.9068926002569e-38\\
0.01	1.28971175009116e-93\\
};
\addlegendentry{$\psi\!=\!20^\circ,v_b\!=\!7$}

\end{axis}

\end{tikzpicture}%
      \vspace{-3em}
      \centerline{\footnotesize{(a) Autocorrelation evaluation}} \medskip
  \end{minipage}
   \begin{minipage}[b]{1\linewidth}
    \centering
\definecolor{mycolor1}{rgb}{0.1811,0.5333,0.3412}%

\begin{tikzpicture}

\begin{axis}[%
width=2.7in,
height=1.8in,
at={(0in,0in)},
scale only axis,
xmode=log,
xmin=0.009,
xmax=45,
xminorticks=true,
xticklabel style = {font=\color{white!15!black},font=\footnotesize},
xlabel style={font=\color{white!15!black},font=\footnotesize},
xlabel={HPBW $\psi$ ($^\circ$)},
ymode=log,
ymin=1e-08,
ymax=0.01,
yminorticks=true,
yticklabel style = {font=\color{white!15!black},font=\footnotesize},
ylabel style={font=\color{white!15!black},font=\footnotesize},
ylabel={Coherence time $T_c$ (s)},
ytick={1e-8,1e-7,1e-6,1e-5,1e-4,1e-3,1e-2},
axis background/.style={fill=white},
xmajorgrids,
xminorgrids,
ymajorgrids,
yminorgrids,
legend style={at={(0,0.74)}, anchor=north west, legend cell align=left, align=left, draw=white!15!black, font=\tiny}
]
\addplot [color=blue, mark=triangle, mark options={solid, blue}, line width=0.8pt, mark size=2.5pt]
  table[row sep=crcr]{%
0.01	0.000858210354325341\\
0.0158489319246111	0.00136999426771655\\
0.0251188643150958	0.00218697465500947\\
0.0398107170553497	0.003307739122992\\
0.0630957344480193	0.0043323022673588\\
0.1	0.00474003174231212\\
0.158489319246111	0.00324876838337023\\
0.251188643150958	0.0019282185207892\\
0.398107170553497	0.00114444190079732\\
0.630957344480193	0.000716916787414796\\
1	0.000457252669896931\\
1.58489319246111	0.000286438407149338\\
2.51188643150958	0.000182691671794092\\
3.98107170553497	0.000116521549170322\\
6.30957344480193	7.43179548783946e-05\\
10	4.91367298592065e-05\\
15.8489319246111	3.42891129357514e-05\\
25.1188643150958	2.6179946215854e-05\\
39.8107170553497	2.18697465500947e-05\\
};
\addlegendentry{$\epsilon=0.3,v_b=0$}

\addplot [color=red, mark=square, mark options={solid, red}, line width=0.8pt, mark size=2.2pt]
  table[row sep=crcr]{%
0.01	0.000655248823736715\\
0.0158489319246111	0.0010273507681793\\
0.0251188643150958	0.00161076153461771\\
0.0398107170553497	0.00248045441431411\\
0.0630957344480193	0.00307808954654567\\
0.1	0.00313396217141822\\
0.158489319246111	0.00176235644057174\\
0.251188643150958	0.00100903504484145\\
0.398107170553497	0.000643566975097729\\
0.630957344480193	0.000381971754928367\\
1	0.00024362325981517\\
1.58489319246111	0.000152613780257896\\
2.51188643150958	9.7337738090392e-05\\
3.98107170553497	6.20824360719125e-05\\
6.30957344480193	3.95964498891879e-05\\
10	2.6179946215854e-05\\
15.8489319246111	1.82691671794092e-05\\
25.1188643150958	1.39486202241965e-05\\
39.8107170553497	1.16521549170322e-05\\
};
\addlegendentry{$\epsilon=0.5,v_b=0$}

\addplot [color=mycolor1, mark=triangle, mark options={solid, rotate=180, mycolor1}, line width=0.8pt, mark size=2.5pt]
  table[row sep=crcr]{%
0.01	0.000465552593115796\\
0.0158489319246111	0.000743179548783946\\
0.0251188643150958	0.00112403866377206\\
0.0398107170553497	0.00149892964872026\\
0.0630957344480193	0.00173093695465906\\
0.1	0.00149892964872026\\
0.158489319246111	0.000729930061435042\\
0.251188643150958	0.000395964498891879\\
0.398107170553497	0.000252547893257738\\
0.630957344480193	0.000149892964872026\\
1	9.56023901095308e-05\\
1.58489319246111	5.98885433491646e-05\\
2.51188643150958	3.75161920154464e-05\\
3.98107170553497	2.39279917342814e-05\\
6.30957344480193	1.55383983127497e-05\\
10	1.0273507681793e-05\\
15.8489319246111	7.16916787414796e-06\\
25.1188643150958	5.37611747455583e-06\\
39.8107170553497	4.57252669896931e-06\\
};
\addlegendentry{$\epsilon=0.7,v_b=0$}

\addplot [color=blue, dashed, mark=triangle, mark options={solid, blue}, line width=0.8pt, mark size=2.5pt]
  table[row sep=crcr]{%
0.01	3.02321302502072e-07\\
0.0158489319246111	3.02321302502072e-07\\
0.0251188643150958	3.02321302502072e-07\\
0.0398107170553497	3.02321302502072e-07\\
0.0630957344480193	3.02321302502072e-07\\
0.1	3.02321302502072e-07\\
0.158489319246111	3.02321302502072e-07\\
0.251188643150958	3.42891129357514e-07\\
0.398107170553497	3.19084898062911e-07\\
0.630957344480193	2.96931484820248e-07\\
1	2.91637757405312e-07\\
1.58489319246111	3.02321302502072e-07\\
2.51188643150958	3.02321302502072e-07\\
3.98107170553497	3.02321302502072e-07\\
6.30957344480193	3.02321302502072e-07\\
10	3.02321302502072e-07\\
15.8489319246111	3.02321302502072e-07\\
25.1188643150958	3.02321302502072e-07\\
39.8107170553497	3.02321302502072e-07\\
};
\addlegendentry{$\epsilon=0.3,v_b=7$}

\addplot [color=red, dashed, mark=square, mark options={solid, red}, line width=0.8pt, mark size=2.2pt]
  table[row sep=crcr]{%
0.01	9.91045856248861e-08\\
0.0158489319246111	9.91045856248861e-08\\
0.0251188643150958	9.91045856248861e-08\\
0.0398107170553497	9.91045856248861e-08\\
0.0630957344480193	9.91045856248861e-08\\
0.1	9.91045856248861e-08\\
0.158489319246111	9.91045856248861e-08\\
0.251188643150958	1.04599895343025e-07\\
0.398107170553497	1.0273507681793e-07\\
0.630957344480193	9.91045856248861e-08\\
1	9.7337738090392e-08\\
1.58489319246111	9.91045856248861e-08\\
2.51188643150958	9.91045856248861e-08\\
3.98107170553497	9.91045856248861e-08\\
6.30957344480193	9.91045856248861e-08\\
10	9.91045856248861e-08\\
15.8489319246111	9.91045856248861e-08\\
25.1188643150958	9.91045856248861e-08\\
39.8107170553497	9.91045856248861e-08\\
};
\addlegendentry{$\epsilon=0.5,v_b=7$}

\addplot [color=mycolor1, dashed, mark=triangle, mark options={solid, rotate=180, mycolor1}, line width=0.8pt, mark size=2.5pt]
  table[row sep=crcr]{%
0.01	2.66551573141612e-08\\
0.0158489319246111	2.66551573141612e-08\\
0.0251188643150958	2.66551573141612e-08\\
0.0398107170553497	2.66551573141612e-08\\
0.0630957344480193	2.66551573141612e-08\\
0.1	2.66551573141612e-08\\
0.158489319246111	2.66551573141612e-08\\
0.251188643150958	2.66551573141612e-08\\
0.398107170553497	2.71389943120822e-08\\
0.630957344480193	2.6179946215854e-08\\
1	2.52547893257738e-08\\
1.58489319246111	2.66551573141612e-08\\
2.51188643150958	2.66551573141612e-08\\
3.98107170553497	2.66551573141612e-08\\
6.30957344480193	2.66551573141612e-08\\
10	2.66551573141612e-08\\
15.8489319246111	2.66551573141612e-08\\
25.1188643150958	2.66551573141612e-08\\
39.8107170553497	2.66551573141612e-08\\
};
\addlegendentry{$\epsilon=0.7,v_b=7$}

\end{axis}
\end{tikzpicture}%
      \vspace{-3em}
      \centerline{\footnotesize{(b) Coherence time evaluation}} \medskip
  \end{minipage}
  \vspace{-1.5em}
  \caption{
    The evaluation of channel autocorrelation and coherence time across various receiver beamwidths.
  }
  \label{fig_sim3}
\end{figure}
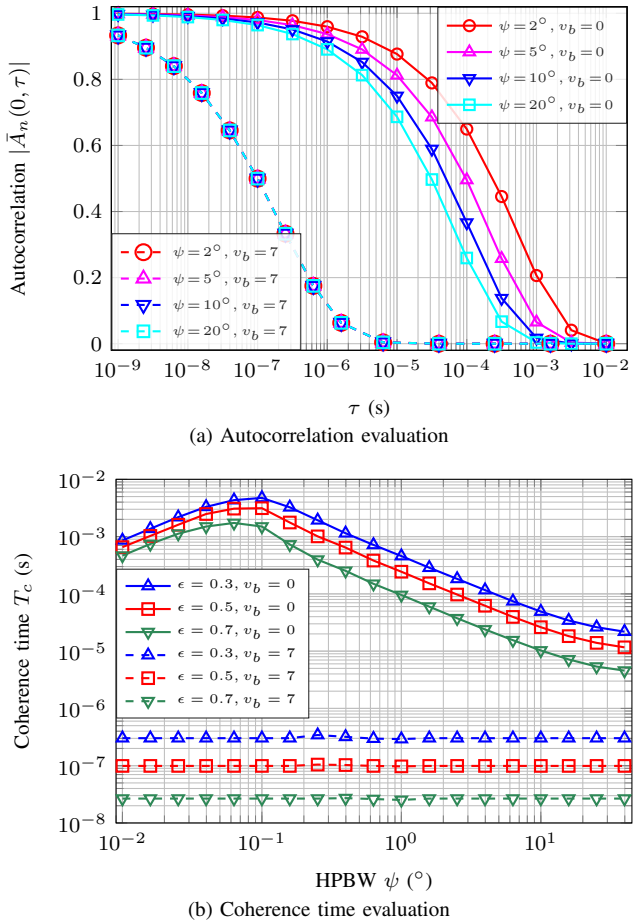

As outlined in~\cite{Va2015Basic} and~\cite{Va2017Impact}, for terrestrial mmWave networks, adjusting receiver beamwidth is an effective strategy to combat the coherence time reduction. We now revisit this issue in the context of non-terrestrial networks. Following~\cite{Va2017Impact}, here we test a scenario where the \ac{los} channel is absent (i.e., $K=0$). Figure~\ref{fig_sim3}-(a) depicts~$|\bar{A}_h(0,\tau)|$ over different \ac{hpbw}~$\psi$ at the \ac{ue}. When the \ac{bs} remains static (solid curves), the channel autocorrelation varies with $\psi$.  However, with the \ac{bs} in motion (dashed curves), adjusting $\psi$ no longer impacts the autocorrelation, as these curves coincide. Figure~\ref{fig_sim3}-(b) provides a more intuitive evaluation showing the relationship between coherence time~$T_c$ and receiver \ac{hpbw}~$\psi$. For static \ac{bs}, there exists an optimal beamwidth~$\psi$ that maximizes~$T_c$, consistent with findings in~\cite{Va2015Basic} and~\cite{Va2017Impact}. As mentioned, this occurs because a narrower beam helps mitigate the impact of destructive multipath components (a positive effect) by focusing power on a subset of scatterers while also exacerbating the misalignment problem (a negative effect). This trade-off leads to the existence of an optimal $\psi$ value that maximizes the channel coherence time. However, with a \unit[7]{km/s} velocity of the \ac{bs}, channel coherence times decrease to extremely low levels and remain almost constant across different $\psi$ values. This is because, in high-mobility scenarios, the misalignment problem becomes significantly more severe. Specifically, the transmit lobe center can easily fail to align with the \ac{ue}, while all \ac{nlos} scatterers only reflect weak signals from the side directions. As a result, the negative impact of multipath components cannot be effectively mitigated. This suggests that optimizing receiver beamwidth no longer extends channel coherence time. These results again unveil that non-terrestrial communications may be infeasible without a \ac{los} link, not only due to weak signal reception but also because of the extremely short channel coherence time.

\section{Conclusion}

This paper studies the coherence time of air-to-ground channels through a statistical analysis of channel autocorrelation. Our analysis reveals that the mobility of the \ac{bs} significantly reduces channel coherence time, posing challenges for reliable communication and beamforming in non-terrestrial networks. However, we find that enhancing the \ac{los} channel can effectively mitigate this effect, leading to improved channel stability. These results underscore, once more, the essential role of the LoS channel in non-terrestrial communications from a channel coherence time perspective.

\bibliography{references}
\bibliographystyle{IEEEtran}

\end{document}